\documentclass{article}

\usepackage{PRIMEarxiv}
\usepackage[utf8]{inputenc}
\usepackage[T1]{fontenc}
\usepackage[hidelinks]{hyperref}
\usepackage{url}
\usepackage{booktabs}
\usepackage{amsfonts}
\usepackage{amsmath}
\usepackage{amssymb}
\usepackage{nicefrac}
\usepackage{microtype}
\usepackage{fancyhdr}
\usepackage{graphicx}
\usepackage{xcolor}
\usepackage{array}
\usepackage{cite}
\usepackage{adjustbox}
\usepackage{subcaption}
\usepackage{placeins}
\usepackage{tikz}
\usetikzlibrary{arrows.meta,positioning,fit,shapes.geometric}
\newif\iftikzavailable
\tikzavailabletrue

\newcommand{\wvds}{\mathrm{WVDS}}
\newcommand{\fft}{\mathrm{FFT}}
\newcommand{\bnd}{\mathrm{BND}}
\newcommand{\raw}{\mathrm{raw}}
\newcommand{\eps}{\varepsilon}
\newcommand{\safeincludegraphics}[2][]{%
\IfFileExists{#2}{\includegraphics[#1]{#2}}{%
\fbox{\begin{minipage}[c][1.2in][c]{0.90\linewidth}\centering
Missing figure file:\\ \texttt{\detokenize{#2}}
\end{minipage}}}}

\title{Early Anomaly-Onset Detection based on Wigner--Ville Distribution Slice Spectra:\\ A Transmission-Grid Test Case}
\author{%
Eduardo Jr Piedad$^{1,2,*}$, 
Eduardo Prieto-Araujo$^{2}$, and Oriol Gomis-Bellmunt$^{2}$\\
$^{1}$DOST--Advanced Science and Technology Institute, Quezon City 1101 Philippines\\
$^{2}$Universitat Polit\`ecnica de Catalunya, 08034 Barcelona, Spain\\
$^{*}$Corresponding author: \texttt{eduardojr.piedad@asti.dost.gov.ph}
}

\begin{document}
\maketitle

\begin{abstract}
Operational disturbance monitoring in power networks requires decisions to be made from waveform windows as they arrive, rather than from completed records after the event. This study evaluates full-vector Wigner--Ville Distribution Slice (WVDS) spectra for sequential anomaly-onset detection in high-voltage grid-voltage waveforms. The approach keeps the bilinear midpoint interaction structure of the Wigner--Ville distribution and represents each 128-sample voltage window by a 128-dimensional slice spectrum, avoiding manually selected fault-frequency markers. WVDS is used with a baseline-normalized deviation (BND) score and is compared against the BND of Fast Fourier Transform (FFT-BND), raw-window autoencoders, FFT autoencoders, and WVDS autoencoders under the same thresholding and three-window persistence rule. A synthetic autoencoder--clustering teacher is used to select RTE fault records that start from an initially normal region and then transition to anomalous behavior. On the filtered test set, FFT-BND achieves the highest sensitivity, whereas WVDS-BND provides the lowest false-alarm operating point, reducing record-level pre-onset false alarms to 0.69\%. The autoencoder comparison follows the same selectivity pattern: WVDS reconstruction decreases false alarms relative to FFT reconstruction but misses more examples. The results indicate that preserved WVD cross-term information can form a selective representation for online grid-waveform anomaly monitoring when false alarms are costly.
\end{abstract}


\section{Introduction}
\label{sec:introduction}

High-voltage transmission-grid disturbances produce transient waveform changes that must be recognized rapidly and reliably. Many machine-learning studies formulate the task as whole-record classification or offline anomaly detection. However, a monitoring system observes a waveform sequentially and must decide when a signal has departed from normal behavior. This is an early anomaly-onset formulation. A detector processes sliding windows in chronological order, raises an alarm only after a fixed persistence rule, and is evaluated by detection delay, missed detections, false alarms before onset, and runtime per window.

The French R\'{e}seau de Transport d'\'{E}lectricit\'{e} (RTE) digital fault-recording database provides measured voltage and current waveforms from high-voltage lines. Its full waveform set, DATA\_S, contains 12,053 fault records, each with six signals $(v_1,v_2,v_3,i_1,i_2,i_3)$ and 21,000 samples at 6400 Hz; the voltage quantization step is 18.310 V and the network nominal frequency is 50 Hz \cite{DatabaseRTE}. A previous autoencoder study on this database used synthetic normal voltage signals, reconstruction error, and clustering for anomaly detection \cite{Piedad2025RTEAutoencoder}. In contrast, the present paper uses the full records to evaluate sequential onset detection and uses the previous autoencoder-clustering approach only as a reference teacher for filtering and onset annotation.

Time--frequency representations are attractive for grid waveforms because transients may redistribute energy across harmonics, interharmonics, and nonstationary components. The Wigner--Ville distribution (WVD) offers high concentration for monocomponent signals but is bilinear and can contain cross terms \cite{Wigner1932,Ville1948,Cohen1995,Boashash2016}. These cross terms are commonly treated as artifacts and are often reduced by smoothing or kernel design. However, their midpoint structure can also encode interactions between dominant and weak waveform components, a behavior that has not been fully explored in literature and for early disturbance detection, hence this study.

This paper formulates grid disturbance analysis as a sequential early-onset
detection task using full waveform records, introduces the full-vector WVDS
spectrum for autoencoder and non-autoencoder detectors, proposes a
baseline-normalized deviation (BND) detector for full
FFT and WVDS vectors, and compares the methods using detection delay,
missed detections, pre-onset false alarms, probability of detection, and
online runtime. The remainder of the paper presents the WVDS representation in Section~\ref{sec:wvds}, the dataset and evaluation protocol in Section~\ref{sec:methodology}, the results in
Section~\ref{sec:results}, and the conclusion in
Section~\ref{sec:conclusion}.

\section{Wigner--Ville Distribution Slice and Cross-Term Geometry}
\label{sec:wvds}

For an analytic signal $z(t)$, the Wigner--Ville distribution (WVD) is
\begin{equation}
    W_z(t,f)
    =
    \int_{-\infty}^{\infty}
    z\left(t+\frac{\tau}{2}\right)
    z^*\left(t-\frac{\tau}{2}\right)
    e^{-j2\pi f\tau}\,d\tau .
    \label{eq:wvd}
\end{equation}

The proposed detector uses a fixed-center WVD extraction referring to this
localized one-dimensional spectrum as a Wigner--Ville Distribution Slice
(WVDS),
\begin{equation}
    M_z(t_c,f)
    =
    \int_{-\infty}^{\infty}
    z\left(t_c+\frac{\tau}{2}\right)
    z^*\left(t_c-\frac{\tau}{2}\right)
    e^{-j2\pi f\tau}\,d\tau .
    \label{eq:wvds_continuous}
\end{equation}
For the $m$-th discrete analysis window $x_m[n]$, $n=0,\ldots,L-1$,
the slice index $n_s=\lfloor(L-1)/2\rfloor$ is used. With samples outside the window
support set to zero, the discrete counterpart implemented in this study is \eqref{eq:wvds_discrete} where $k=0,\ldots,K-1 $.
\begin{equation}
    W_m[k]
    =
    \sum_{\ell=-\Lambda}^{\Lambda}
    x_m[n_s+\ell]x_m^*[n_s-\ell]
    e^{-j2\pi k\ell/K}.
    \label{eq:wvds_discrete}
\end{equation}
where $\Lambda=\min(n_s,L-1-n_s)$ and $K=128$ in the experiments. The full
WVDS vector used by the detector is
\begin{equation}
    \mathbf{z}_{m,\wvds}
    =
    \left[
    |W_m[0]|,|W_m[1]|,\ldots,|W_m[K-1]|
    \right]^\top
    \in\mathbb{R}^{128}.
    \label{eq:wvds_vector}
\end{equation}
The term WVDS emphasizes the fixed-center slice used by the detector; it is
not a new transform separate from the standard WVD. In this paper, each
128-sample sliding window is mapped to the complete 128-dimensional WVDS
spectrum and the complete vector is passed to the detector. No handcrafted
marker selection or fault-frequency bin selection is used in the experimental
comparison.

The usefulness of WVDS follows from its bilinear cross-term geometry. Consider
a multicomponent analytic signal
\begin{equation}
    z(t)=\sum_{q=1}^{Q} A_q e^{j2\pi f_q t},
    \label{eq:multicomponent_signal}
\end{equation}
where $A_q$ is complex. Substitution of
\eqref{eq:multicomponent_signal} into \eqref{eq:wvds_discrete} gives
\begin{equation}
    M_z(t_c,f)
    =
    \sum_{q=1}^{Q}\sum_{r=1}^{Q}
    A_q A_r^*
    e^{j2\pi(f_q-f_r)t_c}
    \delta\left(f-\frac{f_q+f_r}{2}\right).
    \label{eq:multicomponent_wvds}
\end{equation}
The terms with $q=r$ are auto-terms located at the component frequencies
$f_q$, whereas the terms with $q\neq r$ are cross terms located at pairwise
midpoint frequencies $(f_q+f_r)/2$. These cross terms are often regarded as
artifacts in time--frequency analysis. Here, these terms are treated as useful
interaction evidence where a weak component may generate a stronger interaction
with a dominant reference than its own auto-term.

For two real sinusoids with $f_a>f_r>0$,
\begin{equation}
    x(t)=A_r\cos(2\pi f_rt+\phi_r)
        +A_a\cos(2\pi f_at+\phi_a),
    \label{eq:real_two_tone_wvds}
\end{equation}
the component list contains both positive and negative frequencies. The WVDS
interaction markers and the corresponding anomaly-frequency recovery
relations can be written compactly as
\begin{equation}
\begin{aligned}
    f_c^{(+)} &= \frac{f_a+f_r}{2},
    &\qquad
    f_a &= 2f_c^{(+)}-f_r, \\
    f_c^{(-)} &= \frac{f_a-f_r}{2},
    &\qquad
    f_a &= f_r+2f_c^{(-)} .
\end{aligned}
\label{eq:wvds_marker_recovery}
\end{equation}
The paired markers also satisfy
\begin{equation}
    f_a=f_c^{(+)}+f_c^{(-)},\qquad
    f_r=f_c^{(+)}-f_c^{(-)} .
    \label{eq:pair_consistency}
\end{equation}

\begin{figure}[!htbp]
    \centering
    \safeincludegraphics[width=\linewidth]{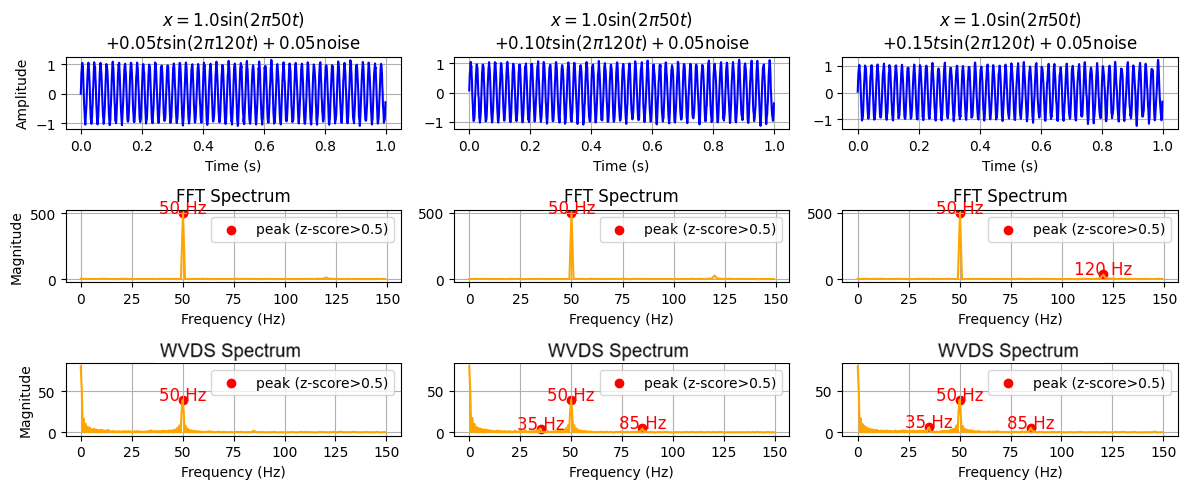}
    \caption{Growing 120-Hz component under a dominant 50-Hz reference. Paired
    WVDS operation produces cross-term markers at 85 Hz and 35 Hz, which both map
    back to the 120-Hz component through $120=2\times85-50$ and
    $120=50+2\times35$.}
    \label{fig:weak_detection}
\end{figure}

Fig.~\ref{fig:weak_detection} illustrates this cross-term mechanism for a
dominant 50-Hz reference and a growing weak 120-Hz component. The resulting
85-Hz and 35-Hz markers are interaction terms rather than physical sinusoids
at those frequencies, and both point back to the same 120-Hz component through
\eqref{eq:wvds_marker_recovery}. Analytic preprocessing suppresses the 35-Hz
marker, whereas real-valued or conjugate-retaining WVDS operation preserves it
as supporting evidence. In the experiments, these markers are not manually
extracted; they remain part of the complete WVDS vector scored by the
detector.

The ideal impulses in \eqref{eq:multicomponent_wvds} are broadened in finite
windows. With lag taper $g(\tau)$ and
$G(\nu)=\int g(\tau)e^{-j2\pi\nu\tau}d\tau$, the windowed WVDS becomes
\begin{equation}
    M_z^{(g)}(t_c,f)
    =
    \sum_{q=1}^{Q}\sum_{r=1}^{Q}
    A_q A_r^*
    e^{j2\pi(f_q-f_r)t_c}
    G\left(f-\frac{f_q+f_r}{2}\right).
    \label{eq:windowed_multicomponent}
\end{equation}
Hence, frequency resolution, leakage, and cross-term separability are
controlled by the lag support and taper. Moreover, the cross-term coefficient
contains the phase factor $e^{j2\pi(f_q-f_r)t_c}$, so its signed value varies
with the slice center. This motivates the use of magnitude, persistence, and
full-vector deviation scoring rather than relying on a single signed
cross-term sample.

For detector input construction, \eqref{eq:wvds_discrete} is evaluated for
each causal sliding window. The full WVDS vector used by the detector is
\begin{equation}
    \mathbf{z}_{m,\wvds}
    =
    \left[
    |M_m[0]|,|M_m[1]|,\ldots,|M_m[K-1]|
    \right]^\top
    \in\mathbb{R}^{128}.
\end{equation}
A log-magnitude transform is applied for numerical stability. Importantly,
all WVDS bins are retained; the cross-term discussion provides physical
interpretation of why the representation can be sensitive to weak waveform
interactions, but the detector itself remains a full-vector detector.

\section{Methodology}
\label{sec:methodology}

\subsection{Dataset and preprocessing}

The voltage phases $v_1$, $v_2$, and $v_3$ from DATA\_S were used as separate one-dimensional record-phase examples. Quantized voltage samples are converted to physical units using the RTE voltage step size. Since $f_s=6400$ Hz and the nominal network frequency is 50 Hz, one cycle contains $L=128$ samples. Windows are advanced with hop $H=32$ samples. Each detector is evaluated causally at the window level, so the alarm time is the end time of the first persistent alarm window.

The public DATA\_S records are fault-event records, and some are already anomalous at the beginning. Therefore the examples that do not contain an initial normal region are removed. Following the previous RTE autoencoder-clustering protocol \cite{Piedad2025RTEAutoencoder}, a synthetic teacher is trained using 1000 synthetic normal voltage windows and validated on 250 synthetic windows. The teacher is a dense autoencoder with architecture $128$--$64$--$32$--$64$--$128$ and a MiniBatchKMeans model with two clusters fitted to the latent codes \cite{Sculley2010}. A Kummerow-style $z=1$ threshold is used to separate normal from low-anomaly behavior \cite{Kummerow2020}. Record-phase examples whose first eight windows are anomalous are removed; retained examples are assigned a teacher-derived onset at the first persistent post-initial anomaly.

This filtering retained 2170 voltage record-phase examples from the full voltage set. Splitting is performed by original RTE record ID, not by phase, so phases from the same physical fault cannot appear in different splits. The final evaluation uses 436 test record-phase examples from 301 unique RTE records.

\begin{figure}[!htbp]
    \centering
    \iftikzavailable
    \resizebox{\linewidth}{!}{%
    \begin{tikzpicture}[
        font=\scriptsize,
        node distance=5mm and 5.5mm,
        block/.style={draw, rounded corners=2pt, align=center,
            minimum height=8.5mm, text width=2.25cm, inner sep=2.4pt},
        wideblock/.style={draw, rounded corners=2pt, align=center,
            minimum height=8.5mm, text width=2.55cm, inner sep=2.4pt},
        decision/.style={diamond, draw, aspect=2.15, align=center,
            inner sep=1pt, text width=1.95cm},
        arrow/.style={-{Latex[length=1.8mm]}, thick},
        group/.style={draw, dashed, rounded corners=3pt, inner sep=4pt}
    ]
        \node[wideblock] (data) {RTE DATA\_S\\voltage phases $v_1,v_2,v_3$\\$21000$ samples, $f_s=6400$ Hz};
        \node[block, right=of data] (window) {Windowing\\$L=128$ samples\\hop $H=32$};
        \node[wideblock, right=of window] (teacher) {Synthetic AE teacher\\1000 train / 250 validation\\AE $128$--$64$--$32$--$64$--$128$\\MiniBatchKMeans, $z=1$};
        \node[decision, right=of teacher] (filter) {first eight\\windows\\normal?};
        \node[block, right=of filter] (onset) {Retain example\\onset: first persistent\\post-initial anomaly};
        \node[block, right=of onset] (split) {Record-ID split\\no phase leakage\\test: 436 phases\\301 records};

        \node[wideblock, below=20mm of split] (train) {Training subset\\normal pre-onset\\windows only};
        \node[wideblock, left=of train] (repr) {Representations\\raw window\\log-FFT magnitude\\log-WVDS magnitude};
        \node[wideblock, left=of repr] (detector) {Detectors\\AE reconstruction MSE\\BND full-vector\\normalized deviation};
        \node[wideblock, left=of detector] (alarm) {Sequential alarm\\validation-normal threshold\\$z\ge1$ and 3 consecutive\\above-threshold windows};
        \node[wideblock, left=of alarm] (metric) {Evaluation\\detection and delay\\pre-onset false alarms\\$P_D(\tau)$ and runtime};

        \draw[arrow] (data) -- (window);
        \draw[arrow] (window) -- (teacher);
        \draw[arrow] (teacher) -- (filter);
        \draw[arrow] (filter) -- node[above, font=\tiny] {yes} (onset);
        \draw[arrow] (onset) -- (split);
        \draw[arrow] (filter.south) -- ++(0,-5mm)
            node[below, font=\tiny, align=center] {no: discard\\already anomalous at start};
        \draw[arrow] (split.south) -- (train.north);
        \draw[arrow] (train) -- (repr);
        \draw[arrow] (repr) -- (detector);
        \draw[arrow] (detector) -- (alarm);
        \draw[arrow] (alarm) -- (metric);

        \node[group, fit=(data)(window)(teacher)(filter)(onset)(split),
            label={[font=\tiny]above:offline record-phase filtering and onset annotation}] {};
        \node[group, fit=(train)(repr)(detector)(alarm)(metric),
            label={[font=\tiny]below:common sequential detector training and evaluation}] {};
    \end{tikzpicture}%
    }
    \fi
    \caption{Sequential onset-detection protocol. The teacher filters voltage
    data and assigns onsets, while detectors train on normal
    pre-onset windows and evaluate the representations under the
    same threshold, persistence rule, and metrics.}
    \label{fig:pipeline}
\end{figure}
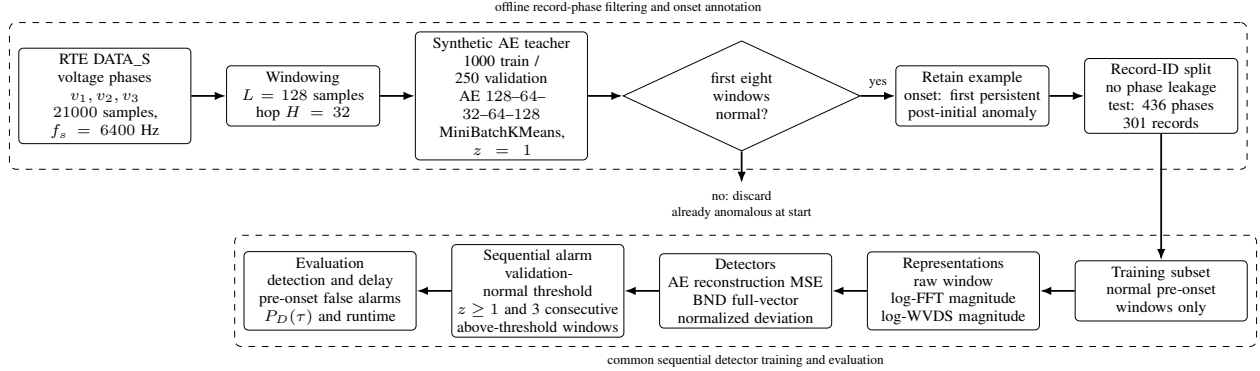

Fig.~\ref{fig:pipeline} summarizes the complete experimental protocol. The
upper path shows the offline teacher stage used for record-phase filtering
and onset annotation. Each RTE voltage phase is converted into overlapping
one-cycle windows, passed through the synthetic AE--MiniBatchKMeans teacher,
and retained only when the initial windows are classified as normal. The
lower path shows the detector-evaluation stage. Only normal pre-onset windows
from the training split are used to estimate detector parameters, while
validation-normal windows set the common threshold. Raw, FFT, and WVDS
representations are then evaluated on the test split using the same
three-window persistence rule and the same detection, delay, false-alarm,
probability-of-detection, and runtime metrics.

\subsection{Compared representations and detectors}
\label{subsec:compared_methods}

Each method receives a 128-dimensional vector so that differences in
performance are attributable to the representation and detector, not to
input dimensionality. Raw-AE uses the original time-domain window. FFT
methods use the full 128-bin FFT magnitude spectrum with the same Hann window
and log-magnitude transform. WVDS methods use the full vector in
\eqref{eq:wvds_vector}. Table~\ref{tab:compared_methods} summarizes the five
methods evaluated under the common sequential onset-detection protocol.

\begin{table}[!htbp]
    \centering
    \caption{Comparison of onset-detection methods.}
    \label{tab:compared_methods}
    \renewcommand{\arraystretch}{1.12}
    \begin{adjustbox}{max width=\linewidth}
    \begin{tabular}{p{0.12\linewidth}p{0.78\linewidth}}
        \toprule
        \textbf{Method} & \textbf{Representation and detector} \\
        \midrule
        Raw-AE &
        Raw 128-sample voltage window scored by autoencoder mean squared
        reconstruction error. \\
        FFT-BND &
        Full 128-bin FFT magnitude vector scored by baseline-normalized
        full-vector deviation. \\
        WVDS-BND &
        Full 128-bin WVDS magnitude vector scored by baseline-normalized
        full-vector deviation. \\
        FFT-AE &
        Full 128-bin FFT magnitude vector scored by autoencoder mean squared
        reconstruction error. \\
        WVDS-AE &
        Full 128-bin WVDS magnitude vector scored by autoencoder mean squared
        reconstruction error. \\
        \bottomrule
    \end{tabular}
    \end{adjustbox}
\end{table}

The BND variants test whether a simple full-vector statistical deviation
score is sufficient for early onset detection, while the AE variants test the
same raw, FFT, and WVDS representations under a learned reconstruction-error
detector. This separation allows two comparisons -- FFT-BND versus WVDS-BND
isolates the effect of the spectral representation under the proposed
non-autoencoder detector, and FFT-AE versus WVDS-AE tests whether the WVDS
representation also benefits an autoencoder-based detector.

For BND, let $q\in\{\fft,\wvds\}$ and let
$\mathbf{z}_{m,q}\in\mathbb{R}^{D}$ be the full vector for a window.
Dimension-wise means and standard deviations are estimated from normal
pre-onset training windows only using
\begin{equation}
    \mu_{q,d}
    =
    \frac{1}{|\mathcal{T}_N|}
    \sum_{(r,m)\in\mathcal{T}_N} z_{m,q,d}^{(r)},
    \qquad
    \sigma_{q,d}
    =
    \operatorname{std}_{(r,m)\in\mathcal{T}_N}
    z_{m,q,d}^{(r)} .
\end{equation}
The normalized deviation and BND anomaly score are defined as 
\begin{equation}
    u_{m,q,d}^{(r)}
    =
    \frac{z_{m,q,d}^{(r)}-\mu_{q,d}}{\sigma_{q,d}+\eps}, \quad
    s_{m,q,\bnd}^{(r)}
    =
    \frac{1}{D}
    \sum_{d=1}^{D}
    \left(u_{m,q,d}^{(r)}\right)^2 .
\label{eq:bnd_score}
\end{equation}
This score uses all vector dimensions and remains interpretable because each
dimension contributes $(u_{m,q,d}^{(r)})^2/D$ to the final anomaly score.

For AE methods, a separate autoencoder is trained for each representation
using the same $128$--$64$--$32$--$64$--$128$ architecture, ReLU hidden
activations, Adam optimizer, batch size 256, and 100 epochs
\cite{Rumelhart1986,Kingma2015Adam}. Each AE is trained only on normal
pre-onset training windows. The anomaly score is the mean squared
reconstruction error of the standardized input vector.

All methods use validation-normal windows to estimate score mean and standard
deviation. A score is above threshold when its validation-normalized
$z$-score is at least 1.0. A valid alarm requires three consecutive
above-threshold windows. The first post-onset valid alarm defines the
detection delay. The detection rate, missed-detection rate, mean and
median delay, record-level and window-level pre-onset false-alarm rates,
probability of detection $P_D(\tau)$, and runtime per window are reported. Runtime includes
representation construction, score computation, thresholding, and alarm-rule
update, and excludes offline training.

\FloatBarrier
\section{Results and Discussion}
\label{sec:results}

\begin{table}[!htbp]
    \centering
    \caption{Detection and online runtime performance on the filtered voltage record-channel test set. Runtime is measured per 128-sample window and excludes offline training. FA denotes pre-onset false alarm.}
    \label{tab:main_results}
    \setlength{\tabcolsep}{2.8pt}
    \renewcommand{\arraystretch}{1.08}
    \begin{adjustbox}{max width=\linewidth}
    \begin{tabular}{lrrrrrrrr}
    \toprule
    Method & Detected & Detection (\%) & Missed (\%) & Mean delay (ms) & Median delay (ms) & Record FA (\%) & Window FA (\%) & Runtime ($\mu$s/window) \\
    \midrule
    Raw-AE & 318/436 & 72.94 & 27.06 & 89.64 $\pm$ 290.62 & \textbf{14.84} & 32.57 & 0.6235 & \textbf{3.180 $\pm$ 0.974} \\
    FFT-BND & \textbf{433/436} & \textbf{99.31} & \textbf{0.69} & \textbf{23.38 $\pm$ 49.97} & 19.84 & 39.45 & 1.0025 & 6.543 $\pm$ 0.460 \\
    WVDS-BND & 384/436 & 88.07 & 11.93 & 54.21 $\pm$ 143.75 & 29.84 & \textbf{0.69} & \textbf{0.0061} & 10.293 $\pm$ 0.556 \\
    FFT-AE & 414/436 & 94.95 & 5.05 & 123.26 $\pm$ 308.31 & 34.84 & 23.17 & 0.7443 & 9.244 $\pm$ 1.516 \\
    WVDS-AE & 377/436 & 86.47 & 13.53 & 106.95 $\pm$ 288.41 & 19.84 & 17.20 & 0.2537 & 13.041 $\pm$ 0.600 \\
    \bottomrule
    \end{tabular}
    \end{adjustbox}
    \end{table}

\begin{table}[!htbp]
    \centering
    \caption{Probability of detection $P_D(\tau)$ at selected post-onset delays, in percent.}
    \label{tab:pd_times}
    \setlength{\tabcolsep}{3pt}
    \begin{adjustbox}{max width=\linewidth}
    \begin{tabular}{lrrrrr}
    \toprule
    Method & 20 ms & 30 ms & 50 ms & 100 ms & 1 s \\
    \midrule
    Raw-AE & 55.28 & 58.49 & 59.86 & 64.68 & 71.56 \\
    FFT-BND & \textbf{78.21} & \textbf{91.97} & \textbf{97.02} & \textbf{98.17} & \textbf{99.31} \\
    WVDS-BND & 11.47 & 63.53 & 80.73 & 84.86 & 87.61 \\
    FFT-AE & 7.34 & 43.81 & 72.48 & 78.44 & 92.43 \\
    WVDS-AE & 44.27 & 54.82 & 61.47 & 70.18 & 84.86 \\
    \bottomrule
    \end{tabular}
    \end{adjustbox}
\end{table}

Table~\ref{tab:main_results} summarizes detection and runtime performance. FFT-BND is the most sensitive method, detecting 433/436 examples with a mean delay of 23.38 ms and a runtime of 6.543 $\mu$s/window. However, this sensitivity comes with a 39.45\% record-level pre-onset false-alarm rate. WVDS-BND detects fewer examples, 384/436, and has a longer median delay of 29.84 ms, but it reduces false alarms sharply by 0.69\% record-level and 0.0061\% window-level false alarms. Hence WVDS provides a conservative low-false-alarm operating point while adding only a small absolute online-runtime cost relative to FFT-BND.

\begin{figure}[!htbp]
    \centering
    \begin{subfigure}[t]{0.48\linewidth}
        \centering
        \safeincludegraphics[width=\linewidth]{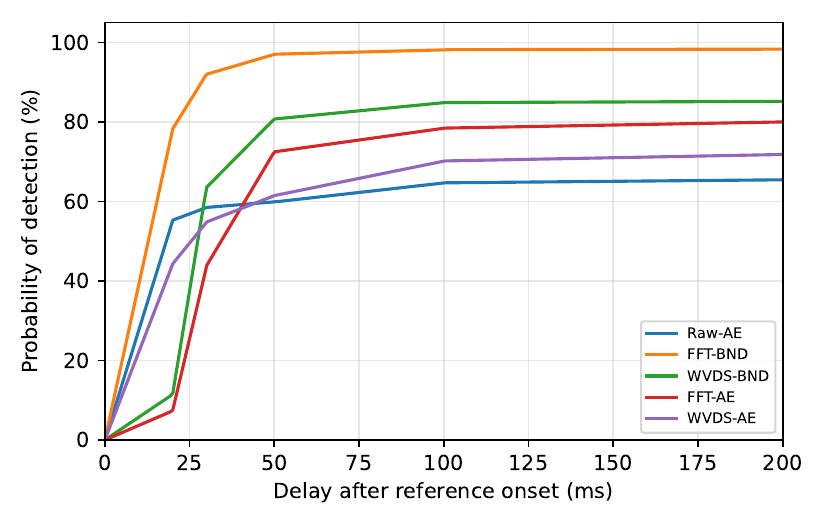}
        \caption{Detection probability versus delay.}
        \label{fig:pd_curve}
    \end{subfigure}\hfill
    \begin{subfigure}[t]{0.48\linewidth}
        \centering
        \safeincludegraphics[width=\linewidth]{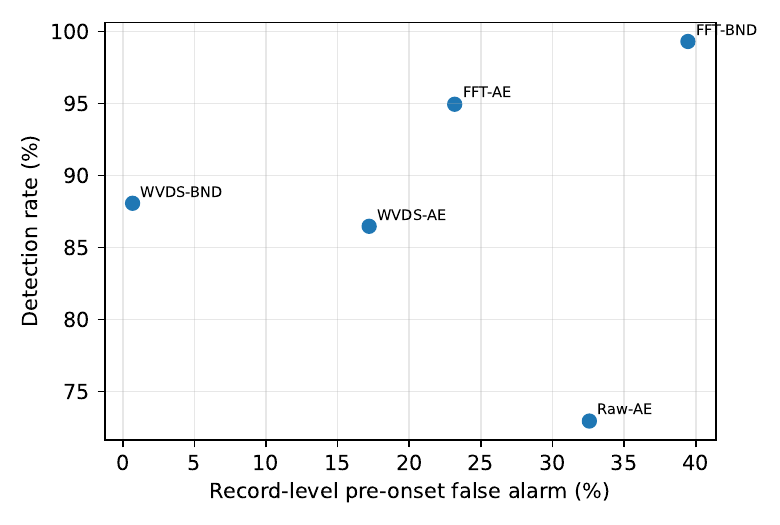}
        \caption{Detection--false-alarm trade-off.}
        \label{fig:fa_tradeoff}
    \end{subfigure}

    \vspace{0.6em}

    \begin{subfigure}[t]{0.48\linewidth}
        \centering
        \safeincludegraphics[width=\linewidth]{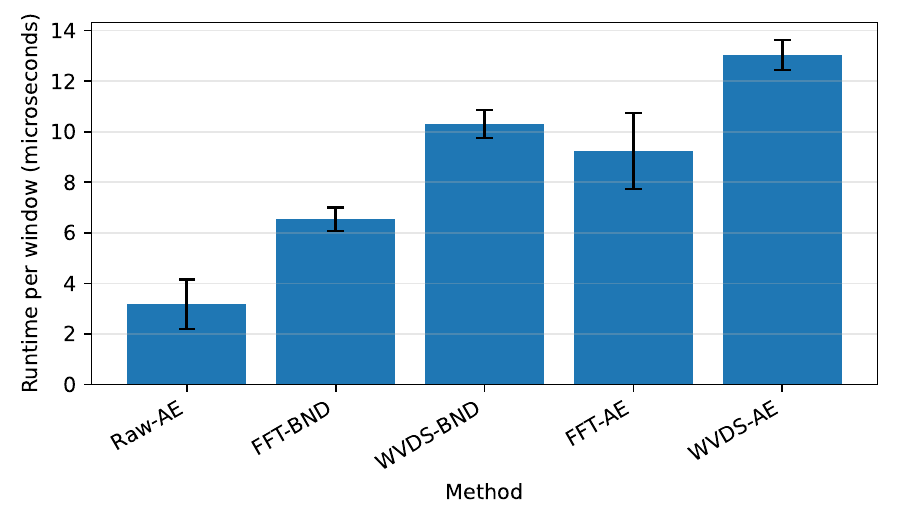}
        \caption{Online runtime per window.}
        \label{fig:runtime}
    \end{subfigure}\hfill
    \begin{subfigure}[t]{0.48\linewidth}
        \centering
        \safeincludegraphics[width=\linewidth]{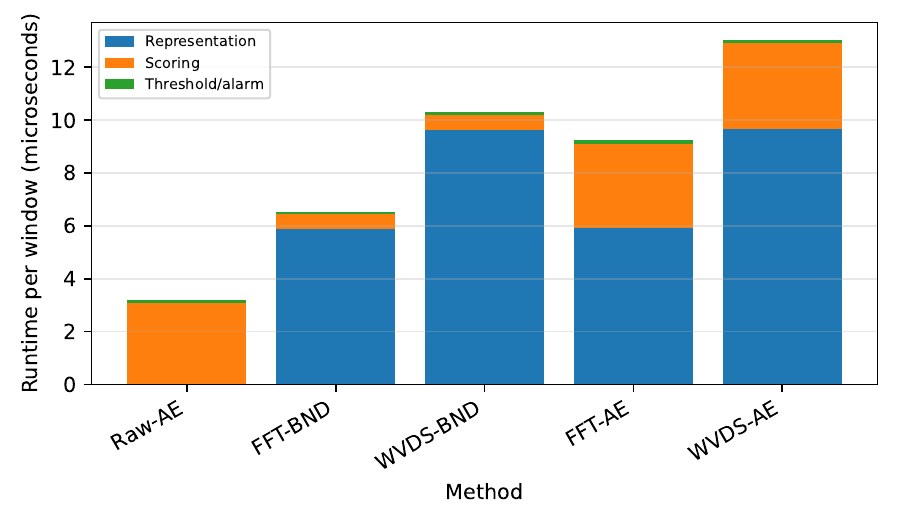}
        \caption{Runtime components.}
        \label{fig:runtime_breakdown}
    \end{subfigure}
    \caption{Summary of detection behavior and online runtime. The plots are arranged in a $2\times2$ layout so that each subfigure remains readable while keeping related results close to the discussion.}
    \label{fig:results_summary}
\end{figure}


\begin{table}[!htbp]
    \centering
    \caption{Runtime components in $\mu$s/window.}
    \label{tab:runtime_components}
    \setlength{\tabcolsep}{15pt}
    \begin{adjustbox}{max width=\linewidth}
    \begin{tabular}{lrrr}
    \toprule
    Method & Representation & Scoring & Alarm \\
    \midrule
    Raw-AE & \textbf{0.004 $\pm$ 0.001} & 3.062 $\pm$ 0.970 & 0.114 $\pm$ 0.011 \\
    FFT-BND & 5.885 $\pm$ 0.425 & \textbf{0.554 $\pm$ 0.055} & \textbf{0.103 $\pm$ 0.019} \\
    WVDS-BND & 9.620 $\pm$ 0.534 & \textbf{0.566 $\pm$ 0.059} & \textbf{0.107 $\pm$ 0.014} \\
    FFT-AE & 5.913 $\pm$ 0.428 & 3.203 $\pm$ 1.320 & 0.127 $\pm$ 0.013 \\
    WVDS-AE & 9.671 $\pm$ 0.504 & 3.239 $\pm$ 0.253 & 0.131 $\pm$ 0.013 \\
    \bottomrule
    \end{tabular}
    \end{adjustbox}
\end{table}

Fig.~\ref{fig:pd_curve} and Table~\ref{tab:pd_times} show early detection probability. FFT-BND dominates early sensitivity, reaching 91.97\% detection by 30 ms and 97.02\% by 50 ms. WVDS-BND reaches 63.53\% at 30 ms and 80.73\% at 50 ms. Therefore, WVDS-BND is not the fastest detector in this configuration; its value is false-alarm suppression and interpretability. This distinction is important for operational monitoring. A permissive detector may alarm earlier, while a conservative detector may be preferred when false trips or alarm fatigue are costly.

Fig.~\ref{fig:fa_tradeoff} visualizes the detection--false-alarm trade-off. WVDS-BND is located near the low-false-alarm region, unlike FFT-BND, FFT-AE, and Raw-AE. Compared with FFT-AE, WVDS-AE has lower record-level false alarms, 17.20\% versus 23.17\%, lower median delay, 19.84 ms versus 34.84 ms, and lower window-level false alarms, but it misses more examples. This suggests that WVDS can improve selectivity for both BND and AE detectors, although the FFT representation remains more sensitive when the threshold is fixed at $z=1$.

Runtime results are shown in Fig.~\ref{fig:runtime} and decomposed in Table~\ref{tab:runtime_components} and Fig.~\ref{fig:runtime_breakdown}. All methods operate in the microsecond-per-window range. Raw-AE is fastest, but its missed-detection rate is 27.06\%. FFT-BND is the best detection--runtime compromise, while WVDS-BND is approximately $10.293/6.543\approx1.57\times$ slower than FFT-BND. The added WVDS cost is dominated by representation construction; BND scoring and alarm logic are nearly identical for FFT and WVDS. Thus, the runtime penalty for WVDS is measurable but small in absolute terms.

The pairwise comparisons clarify the role of WVDS. FFT-BND detected 49 examples that WVDS-BND missed, whereas WVDS-BND detected no examples missed by FFT-BND; among examples detected by both, WVDS-BND was later by a median of 10 ms. Thus, FFT-BND dominated WVDS-BND in sensitivity. The latter, however, provides a substantially lower false-alarm regime. For autoencoders, WVDS-AE was earlier than FFT-AE in 63.11\% of commonly detected cases and had lower false-alarm rates, but FFT-AE detected more examples overall. These results support the interpretation that WVDS emphasizes waveform interaction changes that are selective but not always as sensitive as FFT magnitude changes under the same threshold.

\section{Conclusion and Future Works}
\label{sec:conclusion}

This paper presented a WVDS-based representation for sequential
anomaly-onset detection in high-voltage grid waveforms. By using the complete
WVDS vector rather than selected frequency markers, the detector captures
changes in the time--frequency slice while preserving a direct connection to
the underlying waveform interactions. The results indicate that WVDS-BND
provides a conservative operating point but it is less sensitive than FFT-BND
under the same threshold and persistence rule, but it substantially reduces
pre-onset false alarms. This behavior makes WVDS-BND more suitable for
settings where false alarms are costly, while FFT-BND remains preferable when
maximum detection sensitivity and earlier alarms are the primary concern.

The comparison with autoencoder-based detectors further suggests that WVDS
can improve selectivity beyond the proposed BND score, although the gain in
false-alarm suppression may come with additional missed detections. Thus,
WVDS should be viewed as a selective full-vector time--frequency
representation rather than a universally dominant replacement for FFT-based
features. The
implemented WVDS detector remains practical for window-level inference, but
it is slower than FFT-BND because of the representation-construction stage.
The advantage is that it avoids constructing and storing a full quadratic
time--frequency surface, as would be required by a complete Wigner--Ville
distribution analysis.

Future work will examine adaptive thresholding,
analytic-signal WVDS variants, cross-term contribution mapping, and
validation against externally annotated relay or disturbance-onset labels. Other applications of early onset-detection will be explored.

\IfFileExists{IEEEbib.bst}{\bibliographystyle{IEEEbib}}{\bibliographystyle{IEEEtran}}
\bibliography{references}

\end{document}